\documentclass[journal=jacsat,manuscript=article]{achemso}

\usepackage{chemformula} 
\usepackage[T1]{fontenc} 

\usepackage{bm}
\usepackage{braket}
\renewcommand{\v}[1]{\bm{\mathrm{#1}}}

\usepackage{csquotes}
\MakeOuterQuote{"}

\author{Deepika Gill}
\affiliation[BigPharma]
{Max-Born-Institute for Non-Linear optics, Max-Born Strasse 2A, 12489 Berlin, Germany}

\author{Ruikai Wu}
\affiliation[BigPharma]
{Max-Born-Institute for Non-Linear optics, Max-Born Strasse 2A, 12489 Berlin, Germany}

\author{Peter Elliott}
\affiliation[BigPharma]
{Scientific Computing Department, Science and Technology Facilities Council UK Research and Innovation (STFC-UKRI), Rutherford Appleton Laboratory, Harwell Campus, Didcot OX11 0QX, United Kingdom}

\author{Sangeeta Sharma}
\affiliation[BigPharma]
{Max-Born-Institute for Non-Linear Optics and Short Pulse Spectroscopy, Max-Born Strasse 2A, 12489 Berlin, Germany}
\alsoaffiliation
{Institute for theoretical solid-state physics, Freie Universit\"at Berlin, Arnimallee 14, 14195 Berlin, Germany.}
\alsoaffiliation
{Halle-Berlin-Regensburg Cluster of Excellence CCE}

\author{Sam Shallcross}
\email{shallcross@mbi-berlin.de}
\affiliation[BigPharma]
{Max-Born-Institute for Non-Linear optics, Max-Born Strasse 2A, 12489 Berlin, Germany}

\title[An \textsf{achemso} demo]
{All optical ultrafast pure spin current in the altermagnet Cr$_2$SO}

\abbreviations{IR,NMR,UV}

\begin{document}

\begin{tocentry}

\includegraphics[width=0.8\textwidth]{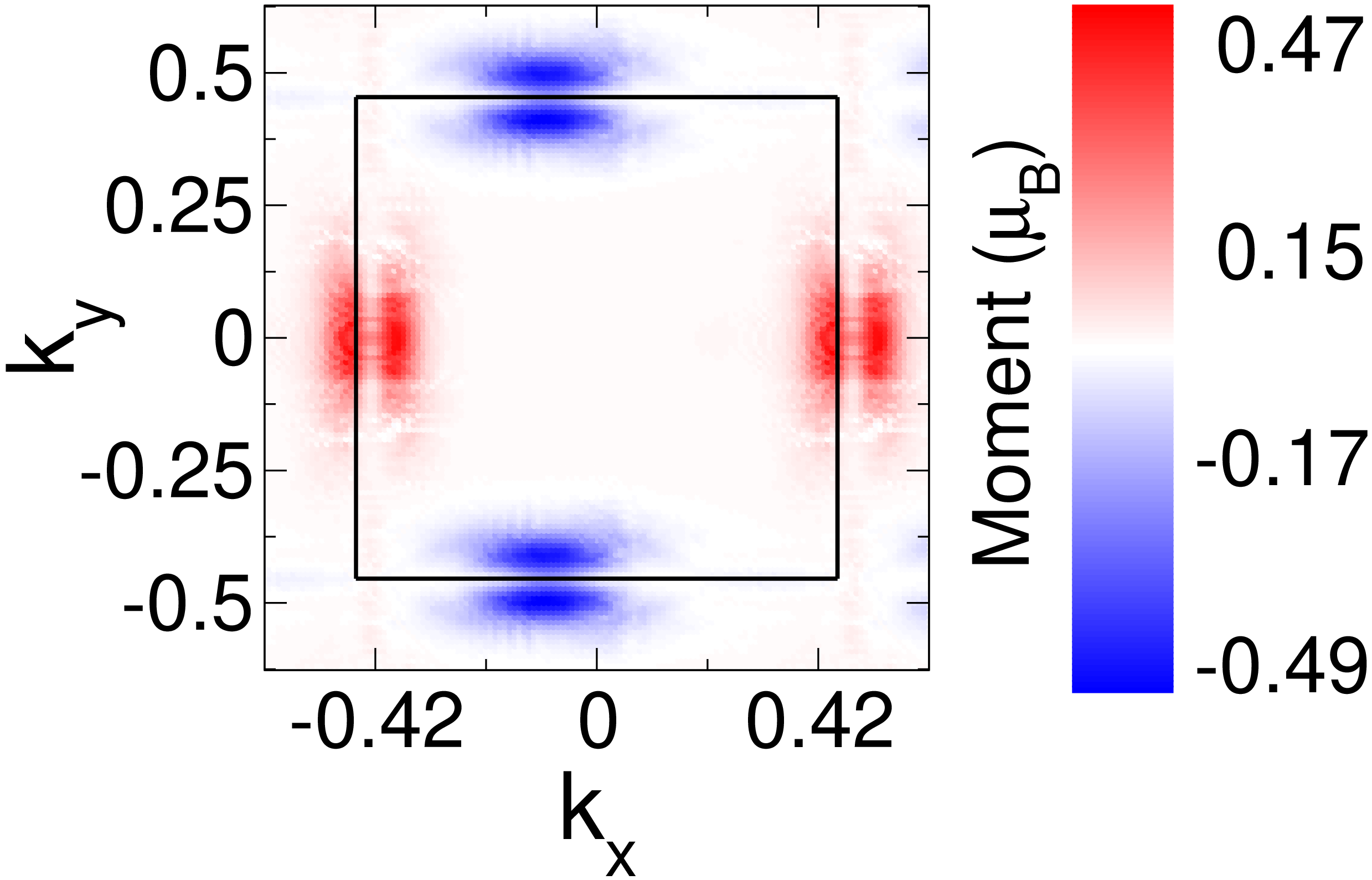}

\end{tocentry}

\begin{abstract}
All-optical generation of pure spin current -- the flow of spin in the absence of a corresponding charge flow -- relies on a symmetry based compensation of valley charge. The 2d $d$-wave altermagnets, ideal spintronics materials due to a very low spin-orbit coupling, possess a magnetic point group and highly anisotropic valley manifolds that would appear to preclude such current compensation, excluding them as materials for the ultrafast generation of pure spin current. Here we show that infra-red valley excitation combined with a THz pulse envelope allows the generation of large and nearly 100\% pure spin currents in the altermagnet Cr$_2$SO. Our approach is based on a valley selection rule coupling linearly polarized light to spin opposite valleys, along with the intrinsic momentum shift that a co-occurring THz pulse imbues a valley spin excitation with. These results thus provide a practical and all-optical route to the generation of pure spin current in $d$-wave 2d altermagnets, opening a route to lightwave control of spin in an environment with very low intrinsic spin mixing.
\end{abstract}

\section{Introduction}

Light-wave control over spin and valley freedoms in the transition metal dichalcogenides (TMDs) is underpinned by a simple physical law: the helicity of circularly polarized light couples selectively to conjugate (opposite momentum) valleys\cite{xiao_coupled_2012}. This elementary fact has birthed a vast range of  phenomena
\cite{mak_control_2012,xiao_nonlinear_2015,
berghauser_inverted_2018,
ishii_optical_2019,
silva_all-optical_2022,sharma_valley_2022,
langer_lightwave_2018}
connected both to the controlling the valley state as an on/off information resource, and to controlling the associated spin- and valley currents. In particular, the TMDs have been proposed as a platform for generating pure spin currents -- the flow of spin in the absence of a corresponding charge flow -- physics that opens new technological horizons of reduced-dissipation transport.

Transition metal dichalcogenides, however, possess a fundamental limitation: they rely on spin-orbit coupling (SOC) to create distinction between electron spins, which ultimately implies “spin mixing” and degradation of any spin signal. In contrast 2d altermagnets
\cite{bai_altermagnetism_2024}
, that share with TMDs the feature of light activated opposite spin valleys, possess spin splitting that is non-relativistic in origin. This prevents deleterious spin mixing -- with concomitant disruption of spin current states -- and in this respect they thus present as ideal candidates for spin- and valleytronics.

Ultrafast\cite{gill_pure_2025} or perturbative
\cite{yu_nonlinear_2014,
shan_optical_2015,xie_photogalvanic_2015,
xie_two-dimensional_2018,chen_photogalvanic_2018,
jin_photoinduced_2018,jiang_robust_2019,
tao_pure_2020,
mu_pure_2021,
xu_pure_2021,zhang_robust_2023,
aftab_recent_2023}
(shift current) generation of a pure spin current relies on compensation between opposite momentum and opposite spin valley charge currents
\cite{yu_nonlinear_2014,shan_optical_2015,
tao_pure_2020,
gill_pure_2025},
and in this respect the mirror symmetry relation between valleys in materials such as the TMDs or bilayer graphene plays a crucial role. The 2d altermagnets, however, possess a very different magnetic point group relation between valleys which, moreover, exhibit highly anisotropic band manifolds. This would appear to preclude the possibility of valley current compensation, and thus to exclude these materials as a platform for the generation of pure spin and valley currents.

Here we propose a remedy to this, based on designing a hybrid light pulse combining a THz envelope with linearly polarized ultrafast valley excitation pulses. Our approach leverages the fact that a THz envelope imbues a non-current carrying valley excitation with current proportional not to the THz electric field $\v E^{THz}$, the intuitive "transport physics" result, but to the THz vector potential amplitude, $\v A^{THz}$. This allows a sequence of ultrafast laser pulses to "sample" the THz waveform at different values of the vector potential, and to thus "tune" of the THz vector potential experienced by each valley excitation so that the post pulse charge current vanishes identically. We show that both for ideal THz monocycles, as well as realistic THz pulse envelopes taken directly from experiment, that this yields ultrafast pure spin and valley currents in $d$-wave 2d altermagnets. Our findings thus open an all-optical route to an altermagnet valleytronics unmolested by spin-orbit coupling.

\section{Spin polarized currents via THz light}

\begin{figure}[t!]
\begin{center}
\includegraphics[width=0.99\textwidth]{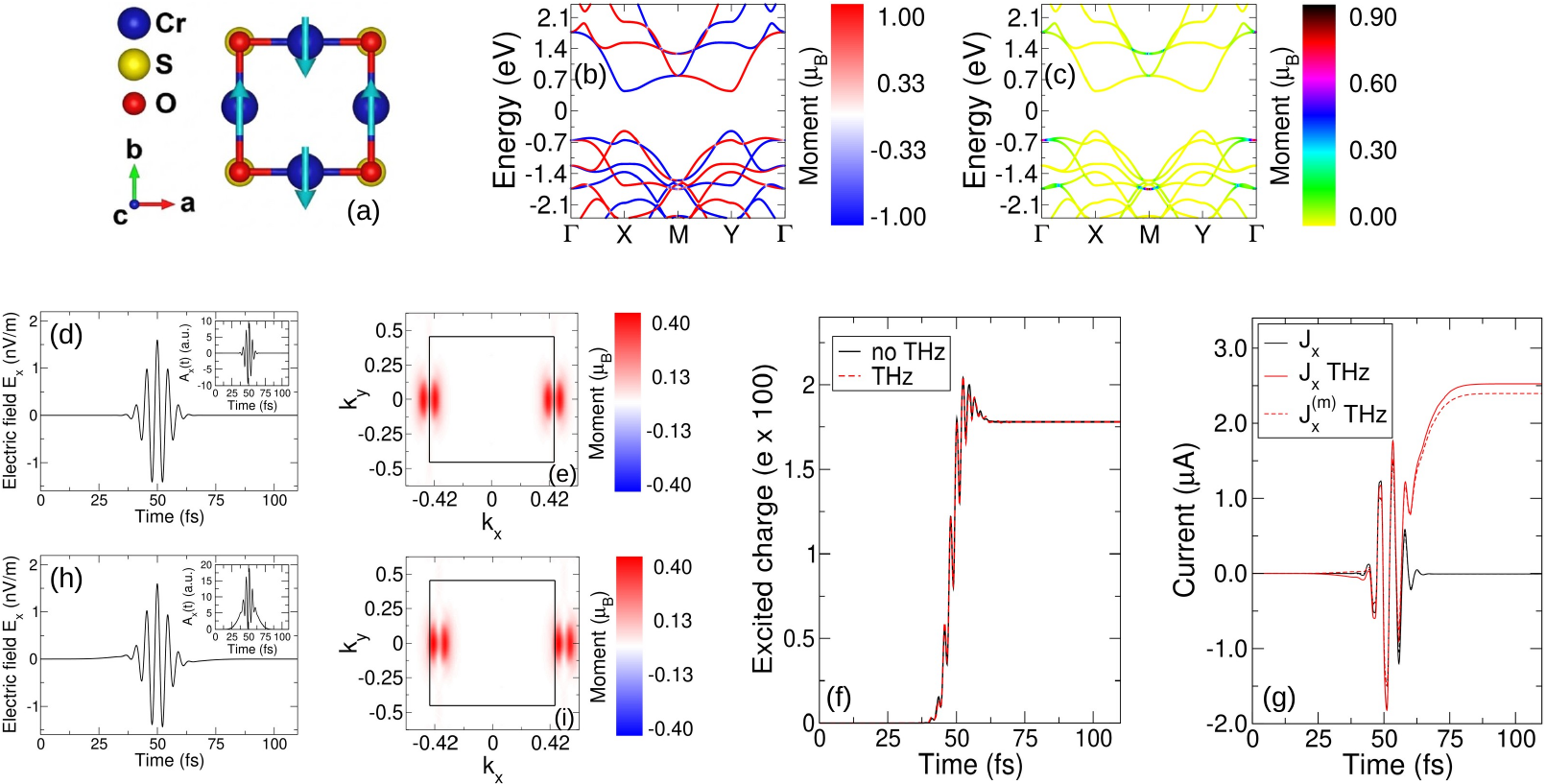}
\caption{{\it Valley excitation and THz generation of spin current in the altermagnet Cr$_2$SO}. (a) The lattice structure of Cr$_2$SO, with anti-ferromagnet alignment of the two Cr sites illustrated. (b) The band structure features opposite spin X and Y valleys with, (c), deviations from collinear magnetic order found only close to the M and $\Gamma$ point. A gap tuned (infra-red) $x$-polarized laser pulse, vector potential shown in (d) with inset displaying the corresponding electric field, generates charge excitation at the X valley, panels (e) and (f) which present respectively the momentum resolved spin excitation post-pulse and the temporal evolution of total Brillouin zone integrated charge, but no post pulse current excitation, panel (g). Augmenting the linearly polarized waveform by a THz envelope, vector potential and corresponding electric field shown in panel (g), dramatically changes this situation. The momentum resolved spin excitation is now clearly displaced from the high symmetry valley centre, and while this renders no difference in the total excited charge, panel (f), it yields significant post pulse spin and charge currents whose near equality (i.e. $\sim$100\% spin polarization) reflects the underlying spin polarized valley bands.
}
\label{fig1}
\end{center}
\end{figure}

As a representative material of the 2d $d$-wave lattice altermagnets we employ Cr$_2$SO
\cite{guo_hidden_2026}, lattice structure shown in Fig.~\ref{fig1}(a). This material features highly localized valley excitation activated by ultrafast linearly polarized laser pulse, without extraneous charge excitation elsewhere in the Brillouin zone. Other similar Lieb lattice materials, e.g. V$_2$Se$_2$O, exhibit unwanted charge excitation at the $\Gamma$ point, spoiling all optical current control. The band structure, Fig.~\ref{fig1}(b,c), reveals a nearly perfectly collinear spin texture, with opposite spin bands at the X and Y point valleys, reflecting the $d$-wave altermagnet exchange splitting.

Linearly polarized laser pulses with polarized in the $x$- or $y$-direction, and with frequency tuned to the infrared gap, excite charge exclusively the X or Y valleys respectively. However, such pulses (unless in the single cycle regime) do not excite current: a multicycle infrared pulse, Fig.~\ref{fig1}(d), excites significant charge at the X valley, visible in the momentum resolved charge excitation shown in panel (e). Note that here and throughout the paper we shown the spin polarized charge excitation. This excited charge -- the temporal evolution of which is shown in panel (f) -- yields however vanishing charge current after the pulse, the temporal evolution of which is presented in panel (g). To calculate the dynamics we employ a tight-binding approach based on a Wannierized {\it ab-initio} band structure; full details of this numerical method can be found in the supplemental document.

Augmenting the linearly polarized infra-red pulse by a THz envelope, Fig.~\ref{fig1}(h),  dramatically changes this situation. The valley excitation is now displaced from the high symmetry valley centre, panel (i), and while the excited charge is nearly identical to the excitation in the absence of the THz envelope, panel (f), the post-pulse valley current is now non-vanishing and large, Fig. (g). This current is nearly 100\% spin polarized, as can be seen by comparison of the charge (full line) and spin (broken line) currents.

The origin of this post-pulse current is the displacement of the excited charge off the valley centre, compare Fig.~\ref{fig1}(e) and (i). In this excited charge distribution current contributions from $+k_x$ and $-k_x$ no longer cancel, as they do for the high symmetry excitation, and thus this lower symmetry distribution possesses a net current.
The dynamical mechanism by which excited charge is displaced from the valley centre can be understood by action of light pulses on a point $\v q$ distant from the valley centre. The infra-red gap tuned pulse acting alone generates no excitation at $\v q$, as here the separation of valence and conduction significantly exceeds that at the valley centre. In combination with a THz envelope, however, excitation can occur via the following dynamical process: (i) the THz frequency cannot excite charge across the infra-red gap, and thus the first half cycle of the THz envelope simply evolves crystal momentum from $\v q \to \v X$; (ii) at X the infra-red pulse can now act to excite charge across the gap; (iii) finally the second half cycle evolves crystal momentum $\v X \to \v q$, and hence the excited charge, at the end of the process, is found at $\v q$ and not the valley centre. Overall, the result of the combined THz and infra-red pulse is to displace the valley excitation off-centre (by an amount $\v A_0^{THz}/c$), as seen in Fig.~\ref{fig1}(i). This so-called "hencomb"\cite{sharma_thz_2023} mechanism has proved useful in controlling spin and charge currents\cite{sharma_giant_2023,
gill_ultrafast_2025}, and their pure counterparts in WSe$_2$, graphene, and bilayer graphene. It also, interestingly, allows for near complete valley polarization of pristine gapless graphene\cite{sharma_combining_2025} -- i.e. valley excitation in the absence of a valley selection rule.

\begin{figure}[t!]
\begin{center}
\includegraphics[width=0.9\textwidth]{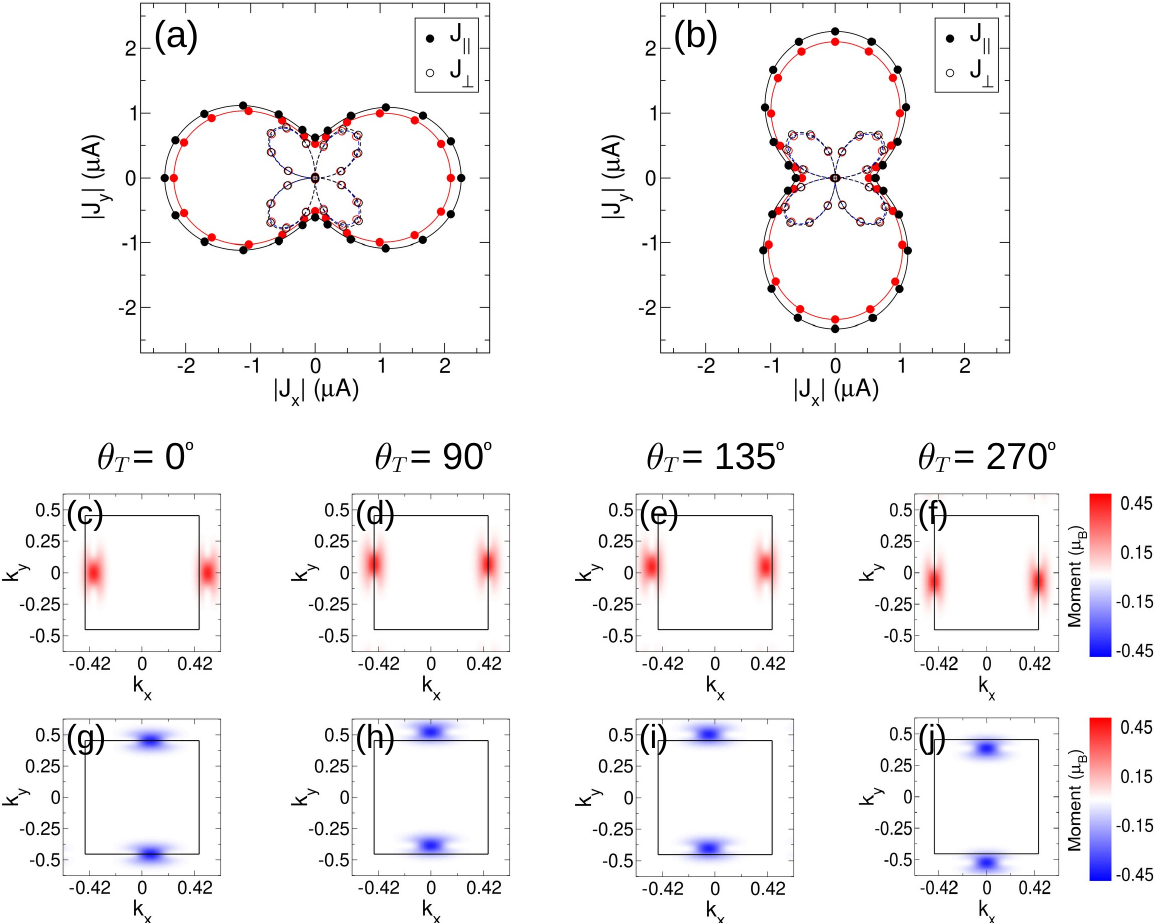}
\caption{{\it Highly anisotropic THz induced  current generation in Cr$_2$SO}. Ultrafast laser excitation of current at the X and Y valleys of the altermagnet Cr$_2$SO generates nearly 100\% pure spin current, but with a highly anisotropic dependence on the in-pane angle of the THz polarization vector, $\theta_T$. (a) The magnitudes of the charge (black) and spin (red) residual current, resolved into components parallel to the THz polarization vector (the full symbols) and perpendicular to the THz polarization vector (the open symbols), shown for the case of excitation at the X valley. For THz polarization aligned with the $x$- or $y$-axis no perpendicular current is generated. (c-f). The corresponding momentum resolved spin excitation for four representative polarization angles. Excitation at the Y valley rotates the current response by $90^\circ$; momentum resolved spin excitations for the same values of THz $\theta_T$ are presented in panels (g-j).
}
\label{fig2}
\end{center}
\end{figure}

This generation of residual current by the displacement of the "native" high symmetry charge excitation of a gap tuned pulse will evidently be very sensitive to the local shape of valley band manifold. Thus in a situation of highly anisotropic valley bands, such as occurs in the 2d altermagnets, a correspondingly anisotropic dependence on the THz polarization angle $\theta_T$ can be expected. That can be immediately discerned in Fig.~\ref{fig2} in which we present, as a polar plot over $\theta_T$, the magnitude of the post-pulse current resolved parallel and perpendicular to the polarization vector: current excitation at $\theta_T = 0^\circ$ is more than double that found for $\theta_T = 90^\circ$. The corresponding momentum resolved charge excitations for four representative values of $\theta_T$ are presented in panels (c-f), revealing the evolving direction of the charge displacement with $\theta_T$. Switching the excitation from the X to the Y valley simply rotates the current response by 90$^\circ$, Fig.~\ref{fig2}(b), with the momentum resolved excitations for the same set of $\theta_T$ presented in panels (g-j). 
One can also note a significant current is induced perpendicular to the THz polarization vector; a feature that again arises from THz induced displacement of valley charge within a anisotropic manifold. 

\section{A THz hybrid pulse for pure spin currents}

\begin{figure}[t!]
\begin{center}
\includegraphics[width=0.99\textwidth]{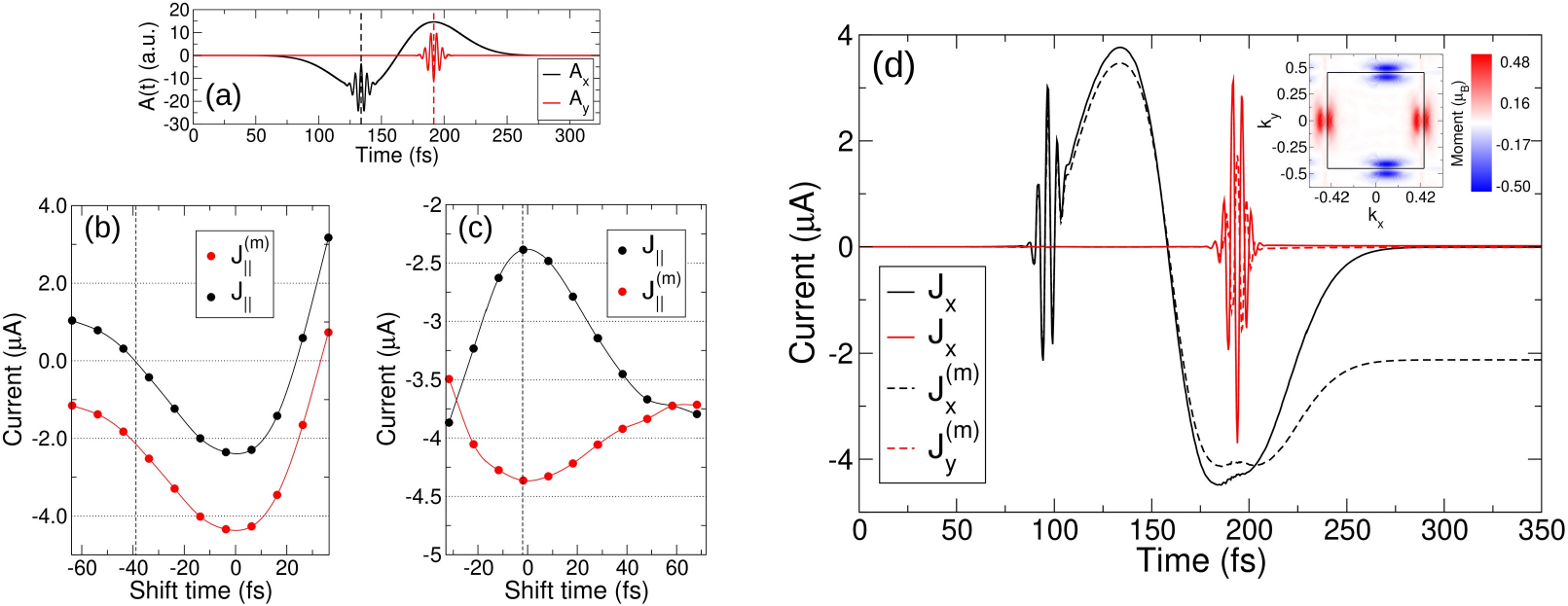}
\caption{{\it Pure spin current generation in Cr$_2$SO}. (a) An ideal THz monocycle is "decorated" by two infra-red valley excitation pulses, with the $x$/$y$-directed polarizations generating excitation at the X and Y valleys respectively. Each of these excitations is imbued with oppositely directed current by the THz envelope, and by shifting the centres of the infra-red pulses within the THz envelope, panels (b) and (c), cancellation of macroscopic charge current ($J$) can be achieved for an X excitation displacement of -39~fs while the spin current ($J^{(m)}$) remains finite. Note that here we show current component parallel to the THz polarization vector (current in the perpendicular is negligible but not identically zero). (d) The temporal evolution of current revealing suppression of charge current occurring after the second excitation pulse, along with the momentum resolved spin excitation post-pulse, shown in the inset panel; the displacements of the X and Y valleys in the excitation density can clearly be discerned. 
}
\label{fig3}
\end{center}
\end{figure}

Such a highly anisotropic valley current response suggests difficulty in cancelling charge currents between valleys: there is no symmetry that ensures this. Schemes based on such cancellation will therefore not work in the context of 2d $d$-wave altermagnets. However the recently proposed THz scheme\cite{gill_ultrafast_2025}
of Gill {\it et al.} is more flexible and, as we  show, can generate nearly 100\% pure spin and valley currents in the altermagnet Cr$_2$SO.

In this approach a THz envelope is "decorated" with two valley excitation pulses, with the net result excitation at both inequivalent valleys each with a displacement given by the THz amplitude at which each excitation occurs. The vector potential of this pulse design is presented in Fig.~\ref{fig3}(a). The X and Y valley currents generated by this pulse will, due to the opposite signs of $\v A^{THz}$ at which the corresponding excitations occur, have opposite sign, and by shifting the pulse centres we can tune for total compensation of the charge current between valleys. Shifting the centre of the X valley excitation (the $x$-directed infra-red pulse which is parallel to the THz polarization vector, allows in this way complete cancellation of charge current, Fig.~\ref{fig3}(b), while retaining finite spin current. The Y valley excitation, generated by the $y$-directed infra-red pulse perpendicular to the THz polarization vector generates significantly less change in current and cannot achieve this compensation, Fig.~\ref{fig3}(b). This reflects that fact if the THz pulse is aligned along $x$ it is perpendicular to the "long axis" of the valley excitation, thereby generating a greater change in current response. The temporal evolution of current through both valley excitations can be seen in panel (d), with the corresponding momentum resolved spin excitation after the pulse, shown as the inset, revealing the different momentum displacements at the X and Y valleys required to achieve charge current compensation.

\begin{figure}[t!]
\begin{center}
\includegraphics[width=0.9\textwidth]{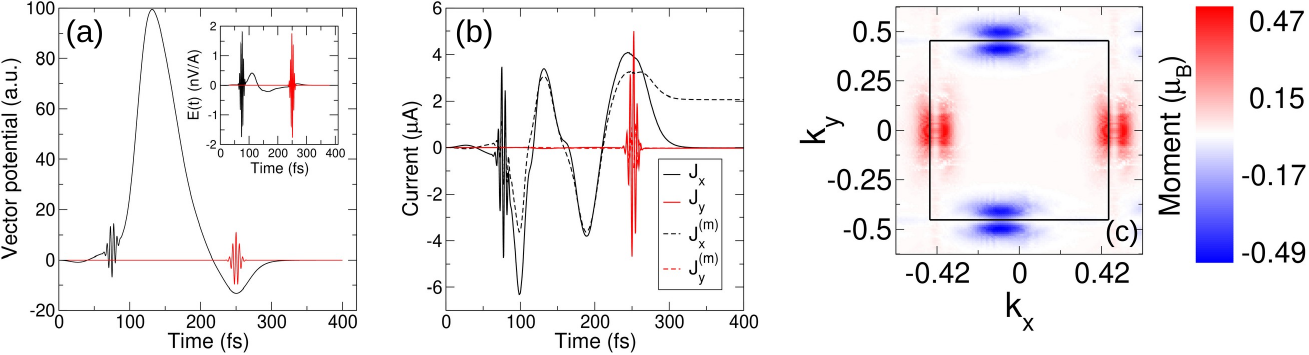}
\caption{{\it Generation of pure spin current utilizing a realistic spintronic emission THz envelope}. (a) An "irregular" THz waveform is decorated by two valley excitation pulses, at the leading edge and in the tail of the pulse respectively; inset panel shows the corresponding electric field. (b) The temporal centres of these pulses ensures momentum shifts of the X and Y valley excitations such that the macroscopic charge current ($J$) vanishes post-pulse while the spin current remains finite ($J^{(m)}$). (c) The momentum resolved spin excitation post-pulse, in which the opposite displacements off the valley centres of the X and Y excitations can be seen.
}
\label{fig4}
\end{center}
\end{figure}

While we have deployed an ideal THz monocycle to illustrate the generation of pure spin current, the scheme adapts to the non-ideal THz envelopes found in experiment. To demonstrate this we take a spintronic emission THz envelope from a recent experiment, and combine this waveform with two valley excitation pulses. The valley compensation tuning process, exactly as described above, then results in excitation pulses fixed to the leading edge of the THz envelope and to the second "minima", as seen in Fig.~\ref{fig4}(a) with the inset presenting the correspond electric field. This generates a similar temporal evolution of current to the ideal THz monocycle, in which the action of the second excitation pulse again results in complete cancellation of charge current, Fig.~\ref{fig4}(b). The momentum resolved excitation, Fig.~\ref{fig4}(c) reveals the greater Y valley excitation displacement, resulting from the fact that this valley has its "long axis" parallel to the THz polarization vector. The scheme we propose thus adapts well to the "irregular" THz envelopes of experiment and, as the precision with which femtosecond scale pulses can be temporally aligned is of the order of a femtosecond, the required precision to generate pure spin current in altermagnets, in a laser setup of combined infra-red and THz pulse generation, is feasible.

\section{Discussion}

The generation of pure spin currents offers the possibility of a classical electronics without wasteful heat dissipation. A significant effort thus acts to investigate schemes in which these currents can be created, with ultrafast light activation approaches requiring on an underlying electronic symmetry. This would appear to exclude the $d$-wave 2d altermagnets as materials for pure spin currents, a disappointing conclusion as, with their very low spin-orbit coupling, these present as ideal candidates for spintronics applications. 

Here we have demonstrated that this difficulty can be overcome by adapting the laser pulse design to the altermagnet valley symmetry and anisotropy: successive excitation of distinct and spin opposite valleys during a co-occurring THz envelope yields perfect charge cancellation but a large (pure) spin current. The underlying mechanism leverages both the coupling of valleys to linearly polarized light pulses that align with the valley centre crystal momentum (and is completely suppressed when the polarization vector is orthogonal to the valley momentum), as well as momentum shift imbued THz light. Our findings thus serve both to illustrate the rich adaptability of "hybrid" pulse designs that combine excitation and momentum shifts, and open a path to exploiting light driven spin currents in an ideal altermagnetic environment of very low intrinsic spin mixing.

\begin{acknowledgement}

Sharma would like to thank SAW for funding through project K612/2024, and Wu would like to thank DFG for funding through project-ID  SH-498-9/1. Gill would like to thank  DFG for funding through project-ID 328545488 TRR227 (project A04). Sharma and Shallcross would like to thank the Leibniz Professorin Program (SAW P118/2021). The authors acknowledge the North-German Supercomputing Alliance (HLRN) for providing HPC resources that have contributed to the research results reported in this paper.

\end{acknowledgement}

\begin{suppinfo}

Supporting information is available:
\begin{itemize}
  \item Filename: SI.pdf -- this document contains supplemental calculations and associated discussion referred to in the text but not elaborated upon.
\end{itemize}

\end{suppinfo}


\begin{mcitethebibliography}{28}
\providecommand*\natexlab[1]{#1}
\providecommand*\mciteSetBstSublistMode[1]{}
\providecommand*\mciteSetBstMaxWidthForm[2]{}
\providecommand*\mciteBstWouldAddEndPuncttrue
  {\def\EndOfBibitem{\unskip.}}
\providecommand*\mciteBstWouldAddEndPunctfalse
  {\let\EndOfBibitem\relax}
\providecommand*\mciteSetBstMidEndSepPunct[3]{}
\providecommand*\mciteSetBstSublistLabelBeginEnd[3]{}
\providecommand*\EndOfBibitem{}
\mciteSetBstSublistMode{f}
\mciteSetBstMaxWidthForm{subitem}{(\alph{mcitesubitemcount})}
\mciteSetBstSublistLabelBeginEnd
  {\mcitemaxwidthsubitemform\space}
  {\relax}
  {\relax}

\bibitem[Xiao \latin{et~al.}(2012)Xiao, Liu, Feng, Xu, and
  Yao]{xiao_coupled_2012}
Xiao,~D.; Liu,~G.-B.; Feng,~W.; Xu,~X.; Yao,~W. Coupled {Spin} and {Valley}
  {Physics} in {Monolayers} of {MoS}$_2$ and {Other} {Group}-{VI}
  {Dichalcogenides}. \emph{Physical Review Letters} \textbf{2012}, \emph{108},
  196802\relax
\mciteBstWouldAddEndPuncttrue
\mciteSetBstMidEndSepPunct{\mcitedefaultmidpunct}
{\mcitedefaultendpunct}{\mcitedefaultseppunct}\relax
\EndOfBibitem
\bibitem[Mak \latin{et~al.}(2012)Mak, He, Shan, and Heinz]{mak_control_2012}
Mak,~K.~F.; He,~K.; Shan,~J.; Heinz,~T.~F. Control of valley polarization in
  monolayer {MoS}$_2$ by optical helicity. \emph{Nature Nanotechnology}
  \textbf{2012}, \emph{7}, 494--498\relax
\mciteBstWouldAddEndPuncttrue
\mciteSetBstMidEndSepPunct{\mcitedefaultmidpunct}
{\mcitedefaultendpunct}{\mcitedefaultseppunct}\relax
\EndOfBibitem
\bibitem[Xiao \latin{et~al.}(2015)Xiao, Ye, Wang, Zhu, Wang, and
  Zhang]{xiao_nonlinear_2015}
Xiao,~J.; Ye,~Z.; Wang,~Y.; Zhu,~H.; Wang,~Y.; Zhang,~X. Nonlinear optical
  selection rule based on valley-exciton locking in monolayer {WS}$_2$.
  \emph{Light: Science \& Applications} \textbf{2015}, \emph{4},
  e366--e366\relax
\mciteBstWouldAddEndPuncttrue
\mciteSetBstMidEndSepPunct{\mcitedefaultmidpunct}
{\mcitedefaultendpunct}{\mcitedefaultseppunct}\relax
\EndOfBibitem
\bibitem[Bergh\"auser \latin{et~al.}(2018)Bergh\"auser, Bernal-Villamil,
  Schmidt, Schneider, Niehues, Erhart, Michaelis~de Vasconcellos, Bratschitsch,
  Knorr, and Malic]{berghauser_inverted_2018}
Bergh\"auser,~G.; Bernal-Villamil,~I.; Schmidt,~R.; Schneider,~R.; Niehues,~I.;
  Erhart,~P.; Michaelis~de Vasconcellos,~S.; Bratschitsch,~R.; Knorr,~A.;
  Malic,~E. Inverted valley polarization in optically excited transition metal
  dichalcogenides. \emph{Nature Communications} \textbf{2018}, \emph{9}, 971,
  Number: 1 Publisher: Nature Publishing Group\relax
\mciteBstWouldAddEndPuncttrue
\mciteSetBstMidEndSepPunct{\mcitedefaultmidpunct}
{\mcitedefaultendpunct}{\mcitedefaultseppunct}\relax
\EndOfBibitem
\bibitem[Ishii \latin{et~al.}(2019)Ishii, Yokoshi, and
  Ishihara]{ishii_optical_2019}
Ishii,~S.; Yokoshi,~N.; Ishihara,~H. Optical selection rule of monolayer
  transition metal dichalcogenide by an optical vortex. \emph{Journal of
  Physics: Conference Series} \textbf{2019}, \emph{1220}, 012056, Publisher:
  IOP Publishing\relax
\mciteBstWouldAddEndPuncttrue
\mciteSetBstMidEndSepPunct{\mcitedefaultmidpunct}
{\mcitedefaultendpunct}{\mcitedefaultseppunct}\relax
\EndOfBibitem
\bibitem[Silva \latin{et~al.}(2022)Silva, Silva, Ivanov, Ivanov, Ivanov,
  Jim{\'e}nez-Gal{\'a}n, and Jim{\'e}nez-Gal{\'a}n]{silva_all-optical_2022}
Silva,~R. E.~F.; Silva,~R. E.~F.; Ivanov,~M.; Ivanov,~M.; Ivanov,~M.;
  Jim{\'e}nez-Gal{\'a}n,~{\'A}.; Jim{\'e}nez-Gal{\'a}n,~{\'A}. All-optical
  valley switch and clock of electronic dephasing. \emph{Optics Express}
  \textbf{2022}, \emph{30}, 30347--30355, Publisher: Optica Publishing
  Group\relax
\mciteBstWouldAddEndPuncttrue
\mciteSetBstMidEndSepPunct{\mcitedefaultmidpunct}
{\mcitedefaultendpunct}{\mcitedefaultseppunct}\relax
\EndOfBibitem
\bibitem[Sharma \latin{et~al.}(2022)Sharma, Elliott, and
  Shallcross]{sharma_valley_2022}
Sharma,~S.; Elliott,~P.; Shallcross,~S. Valley control by linearly polarized
  laser pulses: example of {WSe}$_{\textrm{2}}$. \emph{Optica} \textbf{2022},
  \emph{9}, 947--952, Publisher: Optica Publishing Group\relax
\mciteBstWouldAddEndPuncttrue
\mciteSetBstMidEndSepPunct{\mcitedefaultmidpunct}
{\mcitedefaultendpunct}{\mcitedefaultseppunct}\relax
\EndOfBibitem
\bibitem[Langer \latin{et~al.}(2018)Langer, Schmid, Schlauderer, Gmitra,
  Fabian, Nagler, Schüller, Korn, Hawkins, Steiner, Huttner, Koch, Kira, and
  Huber]{langer_lightwave_2018}
Langer,~F.; Schmid,~C.~P.; Schlauderer,~S.; Gmitra,~M.; Fabian,~J.; Nagler,~P.;
  Schüller,~C.; Korn,~T.; Hawkins,~P.~G.; Steiner,~J.~T.; Huttner,~U.;
  Koch,~S.~W.; Kira,~M.; Huber,~R. Lightwave valleytronics in a monolayer of
  tungsten diselenide. \emph{Nature} \textbf{2018}, \emph{557}, 76--80\relax
\mciteBstWouldAddEndPuncttrue
\mciteSetBstMidEndSepPunct{\mcitedefaultmidpunct}
{\mcitedefaultendpunct}{\mcitedefaultseppunct}\relax
\EndOfBibitem
\bibitem[Bai \latin{et~al.}(2024)Bai, Feng, Liu, Šmejkal, Mokrousov, and
  Yao]{bai_altermagnetism_2024}
Bai,~L.; Feng,~W.; Liu,~S.; Šmejkal,~L.; Mokrousov,~Y.; Yao,~Y.
  Altermagnetism: {Exploring} {New} {Frontiers} in {Magnetism} and
  {Spintronics}. \emph{Advanced Functional Materials} \textbf{2024}, \emph{34},
  2409327\relax
\mciteBstWouldAddEndPuncttrue
\mciteSetBstMidEndSepPunct{\mcitedefaultmidpunct}
{\mcitedefaultendpunct}{\mcitedefaultseppunct}\relax
\EndOfBibitem
\bibitem[Gill \latin{et~al.}(2025)Gill, Sharma, and Shallcross]{gill_pure_2025}
Gill,~D.; Sharma,~S.; Shallcross,~S. Pure {Spin} {Currents} via {Antisymmetric}
  {Light}. \emph{Nano Letters} \textbf{2025}, \emph{25}, 9913--9917, Publisher:
  American Chemical Society\relax
\mciteBstWouldAddEndPuncttrue
\mciteSetBstMidEndSepPunct{\mcitedefaultmidpunct}
{\mcitedefaultendpunct}{\mcitedefaultseppunct}\relax
\EndOfBibitem
\bibitem[Yu \latin{et~al.}(2014)Yu, Wu, Liu, Xu, and Yao]{yu_nonlinear_2014}
Yu,~H.; Wu,~Y.; Liu,~G.-B.; Xu,~X.; Yao,~W. Nonlinear {Valley} and {Spin}
  {Currents} from {Fermi} {Pocket} {Anisotropy} in {2D} {Crystals}.
  \emph{Physical Review Letters} \textbf{2014}, \emph{113}, 156603\relax
\mciteBstWouldAddEndPuncttrue
\mciteSetBstMidEndSepPunct{\mcitedefaultmidpunct}
{\mcitedefaultendpunct}{\mcitedefaultseppunct}\relax
\EndOfBibitem
\bibitem[Shan \latin{et~al.}(2015)Shan, Zhou, and Xiao]{shan_optical_2015}
Shan,~W.-Y.; Zhou,~J.; Xiao,~D. Optical generation and detection of pure valley
  current in monolayer transition-metal dichalcogenides. \emph{Physical Review
  B} \textbf{2015}, \emph{91}, 035402\relax
\mciteBstWouldAddEndPuncttrue
\mciteSetBstMidEndSepPunct{\mcitedefaultmidpunct}
{\mcitedefaultendpunct}{\mcitedefaultseppunct}\relax
\EndOfBibitem
\bibitem[Xie \latin{et~al.}(2015)Xie, Zhang, Zhu, Liu, and
  Guo]{xie_photogalvanic_2015}
Xie,~Y.; Zhang,~L.; Zhu,~Y.; Liu,~L.; Guo,~H. Photogalvanic effect in monolayer
  black phosphorus. \emph{Nanotechnology} \textbf{2015}, \emph{26},
  455202\relax
\mciteBstWouldAddEndPuncttrue
\mciteSetBstMidEndSepPunct{\mcitedefaultmidpunct}
{\mcitedefaultendpunct}{\mcitedefaultseppunct}\relax
\EndOfBibitem
\bibitem[Xie \latin{et~al.}(2018)Xie, Chen, Wu, Hu, Wang, Wang, and
  Guo]{xie_two-dimensional_2018}
Xie,~Y.; Chen,~M.; Wu,~Z.; Hu,~Y.; Wang,~Y.; Wang,~J.; Guo,~H.
  Two-{Dimensional} {Photogalvanic} {Spin}-{Battery}. \emph{Physical Review
  Applied} \textbf{2018}, \emph{10}, 034005, Publisher: American Physical
  Society\relax
\mciteBstWouldAddEndPuncttrue
\mciteSetBstMidEndSepPunct{\mcitedefaultmidpunct}
{\mcitedefaultendpunct}{\mcitedefaultseppunct}\relax
\EndOfBibitem
\bibitem[Chen \latin{et~al.}(2018)Chen, Zhang, Zhang, Zheng, Xiao, Jia, and
  Wang]{chen_photogalvanic_2018}
Chen,~J.; Zhang,~L.; Zhang,~L.; Zheng,~X.; Xiao,~L.; Jia,~S.; Wang,~J.
  Photogalvanic effect induced fully spin polarized current and pure spin
  current in zigzag {SiC} nanoribbons. \emph{Physical Chemistry Chemical
  Physics} \textbf{2018}, \emph{20}, 26744--26751, Publisher: The Royal Society
  of Chemistry\relax
\mciteBstWouldAddEndPuncttrue
\mciteSetBstMidEndSepPunct{\mcitedefaultmidpunct}
{\mcitedefaultendpunct}{\mcitedefaultseppunct}\relax
\EndOfBibitem
\bibitem[Jin \latin{et~al.}(2018)Jin, Li, Wang, and Yu]{jin_photoinduced_2018}
Jin,~H.; Li,~J.; Wang,~T.; Yu,~Y. Photoinduced pure spin-current in
  triangulene-based nano-devices. \emph{Carbon} \textbf{2018}, \emph{137},
  1--5\relax
\mciteBstWouldAddEndPuncttrue
\mciteSetBstMidEndSepPunct{\mcitedefaultmidpunct}
{\mcitedefaultendpunct}{\mcitedefaultseppunct}\relax
\EndOfBibitem
\bibitem[Jiang \latin{et~al.}(2019)Jiang, Kang, Tao, Cao, Hao, Zheng, Zhang,
  and Zeng]{jiang_robust_2019}
Jiang,~P.; Kang,~L.; Tao,~X.; Cao,~N.; Hao,~H.; Zheng,~X.; Zhang,~L.; Zeng,~Z.
  Robust generation of half-metallic transport and pure spin current with
  photogalvanic effect in zigzag silicene nanoribbons. \emph{Journal of
  Physics: Condensed Matter} \textbf{2019}, \emph{31}, 495701, Publisher: IOP
  Publishing\relax
\mciteBstWouldAddEndPuncttrue
\mciteSetBstMidEndSepPunct{\mcitedefaultmidpunct}
{\mcitedefaultendpunct}{\mcitedefaultseppunct}\relax
\EndOfBibitem
\bibitem[Tao \latin{et~al.}(2020)Tao, Jiang, Hao, Zheng, Zhang, and
  Zeng]{tao_pure_2020}
Tao,~X.; Jiang,~P.; Hao,~H.; Zheng,~X.; Zhang,~L.; Zeng,~Z. Pure spin current
  generation via photogalvanic effect with spatial inversion symmetry.
  \emph{Physical Review B} \textbf{2020}, \emph{102}, 081402, Publisher:
  American Physical Society\relax
\mciteBstWouldAddEndPuncttrue
\mciteSetBstMidEndSepPunct{\mcitedefaultmidpunct}
{\mcitedefaultendpunct}{\mcitedefaultseppunct}\relax
\EndOfBibitem
\bibitem[Mu \latin{et~al.}(2021)Mu, Pan, and Zhou]{mu_pure_2021}
Mu,~X.; Pan,~Y.; Zhou,~J. Pure bulk orbital and spin photocurrent in
  two-dimensional ferroelectric materials. \emph{npj Computational Materials}
  \textbf{2021}, \emph{7}, 1--10, Number: 1 Publisher: Nature Publishing
  Group\relax
\mciteBstWouldAddEndPuncttrue
\mciteSetBstMidEndSepPunct{\mcitedefaultmidpunct}
{\mcitedefaultendpunct}{\mcitedefaultseppunct}\relax
\EndOfBibitem
\bibitem[Xu \latin{et~al.}(2021)Xu, Wang, Zhou, and Li]{xu_pure_2021}
Xu,~H.; Wang,~H.; Zhou,~J.; Li,~J. Pure spin photocurrent in
  non-centrosymmetric crystals: bulk spin photovoltaic effect. \emph{Nature
  Communications} \textbf{2021}, \emph{12}, 4330, Number: 1 Publisher: Nature
  Publishing Group\relax
\mciteBstWouldAddEndPuncttrue
\mciteSetBstMidEndSepPunct{\mcitedefaultmidpunct}
{\mcitedefaultendpunct}{\mcitedefaultseppunct}\relax
\EndOfBibitem
\bibitem[Zhang \latin{et~al.}(2023)Zhang, Tao, Li, Liu, Wang, and
  Yin]{zhang_robust_2023}
Zhang,~B.; Tao,~B.; Li,~H.; Liu,~X.; Wang,~Y.; Yin,~H. Robust pure spin
  currents in a binuclear ferric phthalocyanine junction driven by the
  photogalvanic effect. \emph{Journal of Physics D: Applied Physics}
  \textbf{2023}, \emph{56}, 295302\relax
\mciteBstWouldAddEndPuncttrue
\mciteSetBstMidEndSepPunct{\mcitedefaultmidpunct}
{\mcitedefaultendpunct}{\mcitedefaultseppunct}\relax
\EndOfBibitem
\bibitem[Aftab \latin{et~al.}(2023)Aftab, Hegazy, and Iqbal]{aftab_recent_2023}
Aftab,~S.; Hegazy,~H.~H.; Iqbal,~M.~Z. Recent advances in {2D} {TMD} circular
  photo-galvanic effects. \emph{Nanoscale} \textbf{2023}, \emph{15},
  3651--3665, Publisher: The Royal Society of Chemistry\relax
\mciteBstWouldAddEndPuncttrue
\mciteSetBstMidEndSepPunct{\mcitedefaultmidpunct}
{\mcitedefaultendpunct}{\mcitedefaultseppunct}\relax
\EndOfBibitem
\bibitem[Guo(2026)]{guo_hidden_2026}
Guo,~S.-D. Hidden altermagnetism. \emph{Frontiers of Physics} \textbf{2026},
  \emph{21}, 25201, arXiv:2411.13795 [cond-mat]\relax
\mciteBstWouldAddEndPuncttrue
\mciteSetBstMidEndSepPunct{\mcitedefaultmidpunct}
{\mcitedefaultendpunct}{\mcitedefaultseppunct}\relax
\EndOfBibitem
\bibitem[Sharma \latin{et~al.}(2023)Sharma, Elliott, and
  Shallcross]{sharma_thz_2023}
Sharma,~S.; Elliott,~P.; Shallcross,~S. {THz} induced giant spin and valley
  currents. \emph{Science Advances} \textbf{2023}, \emph{9}, eadf3673,
  Publisher: American Association for the Advancement of Science\relax
\mciteBstWouldAddEndPuncttrue
\mciteSetBstMidEndSepPunct{\mcitedefaultmidpunct}
{\mcitedefaultendpunct}{\mcitedefaultseppunct}\relax
\EndOfBibitem
\bibitem[Sharma \latin{et~al.}(2023)Sharma, Gill, and
  Shallcross]{sharma_giant_2023}
Sharma,~S.; Gill,~D.; Shallcross,~S. Giant and {Controllable} {Valley}
  {Currents} in {Graphene} by {Double} {Pumped} {THz} {Light}. \emph{Nano
  Letters} \textbf{2023}, \emph{23}, 10305--10310\relax
\mciteBstWouldAddEndPuncttrue
\mciteSetBstMidEndSepPunct{\mcitedefaultmidpunct}
{\mcitedefaultendpunct}{\mcitedefaultseppunct}\relax
\EndOfBibitem
\bibitem[Gill \latin{et~al.}(2025)Gill, Sharma, Dewhurst, and
  Shallcross]{gill_ultrafast_2025}
Gill,~D.; Sharma,~S.; Dewhurst,~J.~K.; Shallcross,~S. Ultrafast all-optical
  generation of pure spin and valley currents. \emph{npj 2D Materials and
  Applications} \textbf{2025}, \emph{9}, 49, Publisher: Nature Publishing
  Group\relax
\mciteBstWouldAddEndPuncttrue
\mciteSetBstMidEndSepPunct{\mcitedefaultmidpunct}
{\mcitedefaultendpunct}{\mcitedefaultseppunct}\relax
\EndOfBibitem
\bibitem[Sharma \latin{et~al.}(2025)Sharma, Gill, Krishna, Dewhurst, Elliott,
  and Shallcross]{sharma_combining_2025}
Sharma,~S.; Gill,~D.; Krishna,~J.; Dewhurst,~J.~K.; Elliott,~P.; Shallcross,~S.
  Combining {THz} and {Infrared} {Light} to {Control} {Valley} {Charge} and
  {Current} in {Gapless} {Graphene}. \emph{Nano Letters} \textbf{2025},
  \emph{25}, 3791--3798\relax
\mciteBstWouldAddEndPuncttrue
\mciteSetBstMidEndSepPunct{\mcitedefaultmidpunct}
{\mcitedefaultendpunct}{\mcitedefaultseppunct}\relax
\EndOfBibitem
\end{mcitethebibliography}
\providecommand{\latin}[1]{#1}
\makeatletter
\providecommand{\doi}
  {\begingroup\let\do\@makeother\dospecials
  \catcode`\{=1 \catcode`\}=2 \doi@aux}
\providecommand{\doi@aux}[1]{\endgroup\texttt{#1}}
\makeatother
\providecommand*\mcitethebibliography{\thebibliography}
\csname @ifundefined\endcsname{endmcitethebibliography}
  {\let\endmcitethebibliography\endthebibliography}{}

\end{document}